\documentclass[12pt,preprint]{aastex}

\shorttitle{Rotating gas onto a Schwarzschild black hole}
\shortauthors{Huerta \& Mendoza}

\usepackage{amsmath}

\begin{document}

\title{A simple accretion model of a rotating gas sphere onto a
       Schwarzschild black hole}
\author{E. A. Huerta \& S. Mendoza}
\affil{Instituto de Astronom\'{\i}a, Universidad Nacional 
       Aut\'onoma de M\'exico, AP 70-264, Distrito Federal 04510,
       M\'exico}
\email{eahuerta@astroscu.unam.mx,sergio@astroscu.unam.mx}

\begin{abstract}
  We construct a simple accretion model of a rotating gas sphere onto
a Schwarzschild black hole.  We show how to build analytic solutions
in terms of Jacobi elliptic functions.  This construction represents
a general relativistic generalisation of the Newtonian accretion model
first proposed by \citet{ulrich}.  In exactly the same form as it occurs
for the Newtonian case, the flow naturally predicts the existence of an
equatorial rotating accretion disc about the hole.  However, the radius
of the disc increases monotonically without limit as the flow reaches
its minimum allowed angular momentum for this particular model.
\end{abstract}



\section{Introduction}
\label{introduction}

  Steady spherically symmetric accretion onto a central gravitational
potential (e.g.  a star) was first investigated by \citet{bondi52}.
This pioneering work turned out to have many different applications
to astrophysical phenomena (see e.g.  \citet{frank02}), despite of the
fact that it was only constructed for curiosity, rather than a realistic
idea to a particular astrophysical situation \citep{bondi05}.  A general
relativistic generalisation to the work of \citeauthor{bondi52} was made
by \citet{michel72}.  Both models can be seen as astrophysical examples
of transonic flows that naturally occur in the Universe.

  Realistic models of spherical accretion require an extra ingredient
that seems inevitable in many astronomical situations.  This is
so because gas clouds, where compact objects are embedded, have a
certain degree of rotation.  This rotation enables the formation of
an equatorial accretion disc for which gas particles rotate about the
central object.  The first steady accretion model, in which a rotating
gas sphere with infinite extent is accreted to a central object was
first investigated by \citet{ulrich}. In his model, \citeauthor{ulrich}
considered a gas cloud rotating as a rigid body and took no account of
pressure gradients associated to the infalling gas.  In other words,
his analysis is approximately ballistic. For, the the initial specific
angular momentum of an infalling particle is small and heating by
radiation as well as viscosity effects are negligible. In addition,
pressure gradients and internal energy changes along the streamlines of
a supersonic flow provide negligible contributions to the momentum and
energy balances respectively \cite[cf.][]{ulrich,cassen,mendoza}.

  A first order general relativistic approximation of a rotating gas
sphere was made by \citet{beloborodov01}.  In their model, they used
approximate solutions for the integration of the geodesic equation
and their boundary conditions are such that the specific
angular momentum for a single particle \( h \leq 2 r_\text{g} \),
where \( r_\text{g} \) is the Schwarzschild radius. In here and in what
follows we use a system of units for which \( G = c = 1 \), where
\( G \) is the gravitational constant and \( c \) the speed of light.
In this article we show that such a model is not a general relativistic
\citeauthor{ulrich} flow, since its appropriate generalisation must
satisfy the inequality \( h \geq 2 r_\text{g} \).  A pseudo--Newtonian
\citet{paczynsky80} numerical approximation of the extreme hyperbolic \(
h = 2 r_\text{g} \) case was discussed by \citet{lee05}.  We show  that
this pseudo--Newtonian numerical approach differs in a significant way
when compared with the complete general relativistic solution.

  In this article we develop a full general relativistic model of
a rotating gas sphere of infinite extent that accretes matter onto
a centrally symmetric Schwarzschild space--time.  We assume that
heating by radiation and viscosity effects are small so that the flow
can be treated as ideal. Since pressure gradients and internal energy
changes along the streamlines of a supersonic flow provide negligible
contributions to the momentum and energy balances respectively,
the flow is well approximated by ballistic trajectories.  We also
assume that the self--gravity of the accreting gas does not change
the structure of the Schwarzschild space--time.  This is of course
true if the mass of the central object that shapes the space--time
is much greater than the mass of the rotating cloud.  With these
assumptions, we find velocity and particle number density fields as
well as the streamlines of the flow in an exact analytic form using
Jacobi Elliptic functions. The remaining thermodynamic quantities are
easily found by assuming a polytropic flow, for which the pressure
is proportional to a power of the particle number density \citep[see
for e.g.][]{stanyukovich}.  In section~\ref{celestial} we state the
main results from general relativity used to solve the model introduced
in section~\ref{accretion-model}.  We show in section~\ref{convergence}
that the general solution converges to the accretion model considered by
\citet{ulrich} and that for the case of a null value for the specific
angular momentum, the velocity field converges to the one described
by \citet{michel72} for a null value of the pressure gradients on the
fluid. The particular case of a minimum specific angular momentum \(
h = 2 r_\text{g} \) is calculated in section~\ref{ultrarelativistic},
and it is shown that the solutions can be found with the aid of simple
hyperbolic functions.  Finally, in section~\ref{discussion} we discuss
the physical consequences implied by this general relativistic model.

\section{Background in celestial mechanics for general relativity}
\label{celestial}

  The main results from relativistic gravity to be used throughout
the article are stated in this section. The reader is referred  to the
general relativity textbooks by \citet{MTW,chandra,daufields,novi} and
\citet{wald} for further details.

  It is well known that the vacuum Schwarzschild solution
describing  the  final product of gravitational collapse contains
a singularity which is hidden by a horizon.  The solution
corresponding to an exterior gravitational field of static, spherically
symmetric body is given by the Schwarzschild metric:

\begin{equation} \mathrm{d}s^{2} = - \left( 1 - \frac{ 2M } { r }
  \right) \mathrm{d}
     t^{2} + \left( 1 - \frac{ 2M }{ r } \right)^{-1} \mathrm{d} r^{2}+
     r^{2} \mathrm{d} \Omega^{2},
\label{eq.1}
\end{equation}

\noindent where  \( \mathrm{d} \Omega^{2} = \mathrm{d} \theta^{2} +
\sin^{2}\theta \, \mathrm{d}\varphi^{2} \) represents the square of
an angular displacement. The total mass of the Schwarzschild field
is represented by \( M \). The temporal, radial, polar and azimuthal 
coordinates are represented respectively by t, r, \( \theta \) and \(
\varphi \).  In equation~\eqref{eq.1}, we have chosen a signature \(
(-, +,+,+) \) for the metric.  In what follows, Greek indices such as  \(
\alpha \), \( \beta \), etc., are used to denote space--time components,
taking values \( 0,\ 1,\ 2 \) and \( 3 \).

  \citet{birkhoff} showed that it is possible to solve the vacuum
Einstein field equations for a general spherically symmetric space--time,
without the static field assumption.  It follows from his calculations
that the Schwarzschild solution remains the only solution of this more
general space--time.

  The behaviour of light rays and test bodies in the exterior gravitational
field of a spherical body is described by analysing both, timelike and
null geodesics. In order to do that, we first note that the Schwarzschild
metric has a parity reflection symmetry, i.e. the transformation \(
\theta \rightarrow \pi - \theta \) leaves the metric unchanged.
Under these considerations it follows that if the initial position and
tangent vector of a geodesic lies in the equatorial plane \( \theta= \pi /
2 \), then the entire geodesic must lie in that particular plane. Every
geodesic can be  brought to an initially equatorial plane by a rotational
isometry and so, without loss of generality, it is possible to restrict
our attention to the study of equatorial geodesics only.

 In what follows we denote the coordinate basis components by \( x^{\mu}
\) and the tangent vector to a curve by \( u^\alpha = \mathrm{d} x^\alpha /
\mathrm{d}\tau \).  For timelike geodesics the parameter \( \tau\) 
can be made to coincide with the proper
time and for null geodesics it only represents an affine parameter.
Under the above circumstances, the geodesics take the following form 
(cf. \citet{wald}):

\begin{equation}
-  \kappa =  \ g_{\alpha \beta} \, u^{\alpha}  u^{\beta} =- \left(
     1-\frac{ 2M }{r} \right ) \dot{t}^2 + \left( 1 - \frac{ 2M }{ r }
     \right)^{-1} \dot{r}^2+  r^2 \dot{\varphi}^2,
\label{eq.2}
\end{equation}

\noindent where 

\begin{displaymath}
 \kappa :=  \left\{
   \begin{array}{ll}
      1 &  \text{for timelike geodesics,} \\
      0 & \text{for null geodesics.}
   \end{array}\right.  
\end{displaymath}

\noindent In the derivation of the geodesic equation~\eqref{eq.2}, there are
two important constants of motion that must be taken into account. The
first of them is

\begin{equation}
       E := - g_{\alpha \beta} \, \xi^{\alpha} u^{\beta}
         = \left( 1 - \frac{ 2M }               { r }
         \right)\frac{\mathrm{d}t}{\mathrm{d}\tau},
\label{eq.3}
\end{equation}

\noindent where \( \xi^{\alpha}\) represents the static Killing vector
and \(E\) is a constant of motion. For timelike geodesics \(E\) represents the
specific energy of a single particle following a given geodesic,
relative to a static observer at infinity.

  The second constant of motion \( h \) is related to the rotational 
Killing field \( \psi^{\alpha}\) by the following relation:

\begin{equation}
  h := g_{\alpha \beta} \psi^{\alpha} u^{\beta}= r^{2} \sin^{2}\theta
    \frac{\mathrm{d}\varphi}{\mathrm{d}\tau}.
\label{eq.4}
\end{equation}
     
\noindent Since we have chosen \( \theta = \pi /  2 \) without loss of
generalisation,  the previous equation takes the form 

\begin{equation}
  h= r ^{2}  \frac{ \mathrm{d} \varphi }{ \mathrm{d} \tau }.
\label{eq.5}
\end{equation}

\noindent For timelike geodesics \( h \) is the specific angular
momentum. The final equation for the geodesics is found by
direct substitution of equations~\eqref{eq.3} and \eqref{eq.5} into
relation~\eqref{eq.2}. From now on, we restrict the analysis to timelike
geodesics only and so, the equation of motion takes the following form:

\begin{equation}
  \left( \frac{ \mathrm{d}r }{ \mathrm{d} \tau } \right)^2 +  
    \left( 1 - \frac{ 2M }{ r } \right) \left( 1 + \frac{ h^{2} }{ r^{2} }
    \right) = E^{2}.
\label{eq.6}
\end{equation}

  This equation shows that the radial motion of a geodesic is very
similar to that of a unit mass particle of energy \( E^{2} \) in ordinary
one dimensional non-relativistic mechanics. The feature provided by
general relativity in equation~\eqref{eq.6} is that, apart from
a Newtonian gravitational term \( - 2M /  r \) and the centrifugal
barrier \( h^{2} / r^{2} \), there is a new attractive potential term \(
-2 M h^{2} /  r^{3}\), that dominates over the centrifugal barrier for
sufficiently small \(r\).

As it is done in the analysis of the Keplerian orbit for Newtonian
gravity (see for example \citet{daumec}), it is useful to consider \(r\) as
a function of  \( \varphi \) instead of  \( \tau \).  Therefore,
equation~\eqref{eq.6} takes the form

\begin{equation}
  \left( \frac{ \mathrm{d} r }{ \mathrm{d} \varphi } \right)^2 = \frac{ 
    2 M r^3 }{ h^2 } - r^2 + 2 M r + \left( E^2 - 1 \right ) \left( 
    \frac{ r^4 }{ h^2 } \right)
\label{eq.7}.
\end{equation}

  Now, letting \( E^{2}-1 :=2 E_\text{tot} \), where \( E_\text{tot} \)
is the total energy of the particle and \( u = r^{-1}\) equation~\eqref{eq.7}
takes the final form

\begin{equation}
  \left( \frac{ \mathrm{d} u }{\mathrm{d} \varphi } \right)^2 =  
    2 M u^3 - u^2 + \frac{ 2 M u }{ h^2 } + \frac{ 2 E_\text{tot} }{
    h^2 }.  
\label{eq.8}
\end{equation}

Let us define \( u := M v / h^2 \) so that the previous equation simplifies
to 

\begin{equation}
  \left( \frac{ \mathrm{d} v }{ \mathrm{d} \phi } \right)^2 =  
    \alpha v^3 -v^2 + 2 v  + \epsilon, 
\label{eq.9}
\end{equation}
	 
\noindent where

\begin{equation}
  \alpha :=2 \left( \frac{ M }{ h } \right)^2 ,\qquad  \epsilon := \frac{2
  E_\text{tot}  h^2 }{ M^2 }. 
\label{eq.10}
\end{equation}
 
  This equation determines the geometry of the geodesics on the invariant
plane labelled by \( \theta = \pi /  2 \). In fact, this equation
governs the geometry of the orbits described on the invariant plane due
to the fact that  the geometry of the geodesics is determined by the
roots of the  cubic equation

\begin{equation}
  f(v) =  \alpha v^{3} -v^{2} +2v  + \epsilon.
\label{eq.10a}
\end{equation}

  The parameter \( \alpha \) provides the difference between the general
relativistic and the  Newtonian case.  In fact, \( \alpha \rightarrow 0 \)
in the Newtonian limit.  Finally, the eccentricity \( \boldsymbol{e} \)  of
the Newtonian orbit  is related to \( \epsilon \) through the relation

\begin{displaymath}
  \boldsymbol{e}^{2} =1+ \epsilon.
\end{displaymath}
 
\section{Accretion model}
\label{accretion-model}

  The model first proposed by \citet{ulrich} describes a
non--relativistic steady accretion flow that  considers a central
object for which fluid particles fall onto it due to its gravitational
potential.  Their initial angular momentum \( h_\infty \) at infinity is
considered small in such a way that this model is a small perturbation
of \citet{bondi52}'s spherical accretion model. The
specific initial conditions far away from the origin combined with the
assumption that radiative processes and viscosity play no important
role on the flow, imply that the streamlines have a parabolic shape.
When fluid particles arrive at the equator they thermalise their velocity
component normal to the equator.  Since the angular momentum for a
particular fluid particle is conserved, it follows that particles orbit
about the central object once they reach the equator.  The radius \(
r_\text{dN} \) of the Newtonian accretion disc, where particles orbit
about the central object, is given by \citep{ulrich,mendoza}

\begin{equation}
  r_\text{dN} = h_\infty^2 / M.
\label{eq.10b}
\end{equation}

\noindent The velocity field and the density profiles are calculated by
energy and mass conservation arguments.

  We consider now a general relativistic \citeauthor{ulrich} situation in
which rotating fluid particles fall onto a central object that generates a
Schwarzschild space--time.  As described in section~\ref{introduction},
our analysis is well described by a ballistic approximation.
The equation of motion for each fluid particle is thus described by
relation~\eqref{eq.9}.

  In order to get quantitative results it is important to establish
the boundary conditions at infinity. The angular momentum is given by
equation~\eqref{eq.5}, so if a particle that falls onto the black hole has
an initial velocity \( v_{0} \) at an initial polar angle \( \theta_{0}
\) and the radial distance between the particle and the black hole is  \(
r_{0} \), then the angular momentum  is given by

\begin{equation}
  h_{\infty * } = r_0^2  \frac{ \mathrm{d} \varphi }{ \mathrm{d} \tau } = 
    r_0\,  \gamma_0\, v_0\, \sin\theta_0,
\label{eq.11}
\end{equation}

\noindent where \( \gamma_0:= \left( 1- v_{0} ^{2} \right)^{-1/2} \)
is the Lorentz factor for the velocity \( v_0 \).  

\noindent In addition, \(h_{\infty *}\) is related to the angular momentum
\( h \) perpendicular to the invariant plane through the relation 

\begin{equation}
   h = h_{\infty *} \sin\theta_{0}.
\label{eq.12a}
\end{equation}

\noindent In the Newtonian case, the specific angular momentum \(h_{\infty
*}\) converges to the value calculated by \citet{ulrich}.

  With the above relations  it is then possible to calculate the
equation for a given fluid particle falling onto the central object.
First of all, equation~\eqref{eq.10a} states that,  if \(f(v)\) is a
cubic polynomial in \( v \), then either all of its roots are real or
one of them is real and the two remaining are a complex--conjugate pair.
The fact that the particle's energy is insufficient to permit its escape
from the black hole's gravitational field requires that \( \epsilon <
0\). This implies that the roots \( v_1,\ v_2,\ \) and \( v_3 \) of \(
f(v) \) are all real and satisfy the inequality  \( v_{1} < v_{2} <
v_{3} \). Thus, \( f(v) \) can be written as

\begin{equation}
  f(v) = \alpha \left( v - v_1 \right) \left( v_2 - v \right)
    \left( v_{3} - v \right).
\label{eq.14}
\end{equation}

  Direct substitution of this relation in equation~\eqref{eq.9} yields the
integration

\begin{equation}
  - \frac{ 2 }{ \left[ \left( v_2 - v_1 \right) \left( v_3 - v_1 \right) 
    \right]^{1/2} }  \int \frac{ \mathrm{d} w }{ \left[ \left( w^2 - w_1^2
    \right) \left( w^{2} - w_{2} ^{2} \right) \right]^{1/2} } =
    \alpha^{1/2}\varphi,
\label{eq.16}
\end{equation}

\noindent where \( w_1^2 = 1 / \left( v_{2} - v_{1} \right) \), 
\( w_{2} ^{2} = 1 / \left( v_{3} - v_{1} \right) \) and \( v =
v_{1} + w^{-2} \). This elliptic integral can be calculated in terms of
Jacobi elliptic functions (see for example \citet{cayley,hancock})
yielding the following result

\begin{equation}
  \frac{ 1 }{ \left( v_{3} - v_{1} \right)^{ 1/2 } } 
    \text{ns}^{-1} \left\{ w \left( v_{2} - v_{1} \right)^{1/2} 
    \right\} = \alpha^{ 1/2 } \varphi.
\label{eq.17}
\end{equation}
 
\noindent  The modulus \( k \) of the Jacobi elliptic function for this
particular problem is given by

\begin{equation} 
   k^{2}= \frac{ v_{2} - v_{1} }{ v_{3} - v_{1} }. 
\label{eq.17a}
\end{equation}

\noindent With the aid of relation~\eqref{eq.17}, the equation of 
the orbit is now obtained:

\begin{equation}
  v  =v_{1} + \left( v_{2} - v_{1} \right) \ \text{sn}^{2} \Big\{ \frac{
    \varphi }{ 2 } \left[ \alpha \left( v_{3} - v_{1} \right) \right]^{
    1/2 } \Big\}.
\label{eq.18}
\end{equation}

\noindent This is a general equation for the orbit. For the particular
case we are interested in, it must resemble the orbit proposed by
Ulrich when  \( \alpha = 0\). Thus, the equation of the orbit must
converge to a parabola in this limit. This is possible if and only if the
eccentricity   \(\boldsymbol{e} = 1\), which in turn implies \( \epsilon =
0\). All these conditions mean that the roots of equation~\eqref{eq.10}
are given by

\begin{equation}
  v_{1} = 0, \quad v_{2} = \frac { 1 - \left( 1 - 8 \alpha \right)^{ 1/2 }
    }{ 2 \alpha }, \quad v_{3} = \frac{ 1 + \left( 1 - 8 \alpha \right)^{ 1/2 
    } }{ 2 \alpha },
\label{eq.18a}
\end{equation}

\noindent and so, the modulus \( k \) of the Jacobi elliptic functions in 
equation~\eqref{eq.17a} takes the form

\begin{equation}
   k^2 = \frac{ 1 - ( 1 - 8 \alpha )^{ 1/2 } }{ 1 + ( 1 - 8 \alpha )^{
     1/2 } }.
\label{eq.18b}
\end{equation}

\noindent Note that the previous equations restrict the value of \( \alpha
\) in such a way that 

\begin{equation}
  0 \leq \alpha \leq 1/8.
\label{eq.18ba}
\end{equation}

\noindent When \( \alpha = 0  \), Ulrich solutions are obtained and the
case \( \alpha = 1 / 8 \)  corresponds to the case
for which the angular momentum \( h = 4 M = 2 r_\text{g} \) reaches a
minimum value.

  The orbit followed by a single particle falling onto a 
Schwarzschild black hole with the Ulrich prescription is then given by

\begin{gather} 
  v = \frac{ p }{ r }  = v_{2} \,  \text{sn}^{2}
  \varphi \beta,
					\label{eq.21}  \\
  \intertext{where}
  \beta := \frac{ \left( \alpha v_{3} \right)^{ 1/2 } }{ 2 } = \left(
    \frac{ 1 + \left( 1 - 8 \alpha \right)^{ 1/2 } }{ 8 }\right)^{ 1/2 },
    \quad p:= \frac{ h^{2} }{ M } = \frac{ h^{2}_{\infty
    *} }{ M } \sin^{2} \theta_0 = r_* \sin^{2} \theta_0,
    					\label{eq.21a}
\end{gather}

\noindent and \(p\) is the \emph{latus rectum} of the generalised
conic. Note that in the Newtonian limit, the length \(r_*\) defined by
equation~\eqref{eq.21a} converges to the radius of the Newtonian disc
\( r_{\text{dN}} \) as shown by relation~\eqref{eq.10b}.

  Before using the equation of the orbit to find out the velocity
field and the particle number density, it is useful to mention
some important properties of the Jacobi elliptic functions, such as
\citep{cayley,hancock}

\begin{gather}
  \text{sn}^{2}(z,k) + \text{cn}^{2}(z,k) = 1
  						\notag \\
  \text{sn}(z,k) \rightarrow  \sin(z), \quad 
    \text{cn}(z,k) \rightarrow \cos(z), \quad 
    \text{dn}(z,k)  \rightarrow 1, \quad
    \text{as}  \quad k \rightarrow 0, 
    						\notag \\
  \frac{ \mathrm{d}}{\mathrm{d}z} \text{sn}(z,k) =  \text{cn}(z,k) \
    \text{dn}(z,k),  \qquad 
    \frac{\mathrm{d}}{\mathrm{d}z} \text{cn}(z,k) = -\text{sn}(z,k) \
    \text{dn}(z,k).
						\label{eq.22t}
\end{gather}

  The relativistic conic equation is obtained by direct substitution
of these relations onto equation~\eqref{eq.21}, giving 

\begin{equation}
   r = \frac{ p }{ v_{2} \left( 1- \text{cn}^{2}\varphi \beta \right) }
\label{eq.22}.
\end{equation}

\noindent This orbit lies on the invariant plane \( \theta = \pi /
2 \).  We now obtain an equation of motion in terms of the polar
coordinate \( \theta \) and the initial polar angle \( \theta_{0}\)
made by a particle when it starts falling onto the black hole.  To do so,
we note the fact that in order to recover  the geometry of the
spherical~3D space as \(\alpha \rightarrow 0 \) it should be fulfilled
that\footnote{\citet{ulrich} showed that \( \cos \varphi = \cos \theta
/ \cos \theta_0  \) using geometrical arguments.  For the general
relativistic limit, one is tempted to generalise this result to \(
\text{cn} \varphi \beta = \text{cn} \theta \beta / \text{cn} \theta_0
\beta \).  However, this very simple analogy does not reproduce the
velocity and particle number density fields for the Newtonian limit.}

\begin{equation}
  \text{cn}^{2} \varphi \beta = \frac { \text{cn}^{2} \theta_{0} \beta +
    \text{cn}^{2} \theta \beta - 1 }{ 2 \text{cn}^{2} \theta_{0} \beta 
    - 1 }.
\label{eq.23}
\end{equation}

  Since the invariant plane passes through the origin of
coordinates, then  the radial coordinate  \( r \) remains the same if
another plane is taken instead of the invariant one. Therefore, the angle  \(
\theta_{0}\) is the same as the one related to the value of the angular
momentum of the particle at infinity (cf. equation~\eqref{eq.11}). Thus,
the equation of the orbit is found  by direct substitution of
equations~\eqref{eq.11} and \eqref{eq.23} into \eqref{eq.22}, and is
given by

\begin{equation}
  r = \frac{ r_*  \sin^{2}\theta_0 \left( 2 \, \text{cn}^{2} \theta_{0} 
    \beta -1 \right) }{ v_{2} \left( \text{cn}^{2}\theta_{0} \beta-
    \text{cn}^{2}\theta \beta \right) }.
\label{eq.24}
\end{equation}
   
\noindent In order to work with dimensionless variables, let us make the
following transformations

\begin{displaymath}
  \frac{ r }{ r_* } \rightarrow r, \qquad
     \frac{ v_{i} }{ v_\text{k} } \rightarrow v_{i} \qquad 
     (i = r,  \ \theta,\ \varphi), \qquad
     \frac{ n }{ n_{0} } \rightarrow  n,  
\end{displaymath}
 
\noindent where
 
\begin{displaymath}
  v_{i} :=  \frac{ \mathrm{d} x^i }{ \mathrm{d}\tau }, \qquad 
    n_{0} := \frac{ \dot{ M } }{ 4 \pi v_\text{k} r^{2}_* }, \qquad 
    v_\text{k} := \left( \frac{ M }{ r_* } \right)^{ 1/2 }. 
\end{displaymath}

\noindent  In the previous relations, the mass accretion rate onto the
black hole is represented by \( \dot M \). The  velocity \( v_\text{k} \)
converges to the Keplerian velocity of a single particle orbiting
about the central object in a circular orbit when \( \alpha = 0 \).
The particle number density \( n_{0} \) converges to the one calculated
by \citet{bondi52} in the Newtonian limit for the same null value of \(
\alpha \).

  Under the above considerations, the equations for the streamlines 
\( r(\theta) \), the velocity field \( v_r,\ v_\theta, v_\varphi \) and the
proper particle number density \( n \) are given by

\begin{gather}
    r = \frac{ \sin^2\theta_0 \left( 2 \text{cn}^{2}\theta_{0} \beta
     -1 \right) }{ v_{2 } \left( \text{cn}^{2}\theta_{0} \beta -
     \text{cn}^{2}\theta \beta \right) },
					\label{eq.25} \\
  v_r = -2 r^{-1/2} \beta \ \frac{ \text{cn} \beta\theta \
    \text{sn}\beta\theta \ \text{dn}\beta\theta }{ \sin\theta } \ f^{
    1/2 }_1 \left( \theta,\, \theta_{0},\, v_{2},\, \beta \right),
					\label{eq.28}  \\
  v_{\theta} = r^{ -1/2 } \frac{ \text{cn}^{2}\theta_{0} \beta -
   \text{cn}^{2}\theta \beta }{ \sin\theta } \ f^{ 1/2 }_{1} \left( 
   \theta, \theta_{0}, v_{2}, \beta \right),
					\label{eq.27} \\
  v_{\varphi} = r^{ -1/2 } \frac{ \sin\theta_{0} }{ \sin\theta } \left(
    \frac{v_{2} \ \left( \text{cn}^{2} \theta_{0} \beta- \text{cn}^{2}
    \theta \beta \right) }{ 2 \text{cn}^2 \theta_{0} \beta -1 } \right)^{
    1/2 },
					\label{eq.26} \\
  n = \frac { r^{ -3/2 } \sin\theta_{0} }{ 2 f^{ 1/2 }_1 \left( \theta,\,
  \theta_{0},\, v_{2},\, \beta \right) \ f_{2} \left( \theta,\, 
  \theta_{0},\, v_{2},\, \beta \right) },
					\label{eq.28a}
\end{gather}

\noindent  where the functions \(f_{1}\) and \(f_{2}\)
are defined by the following relations:

\begin{gather*}
  f_{1} \left( \theta, \theta_{0}, v_{2}, \beta \right) := 
    \frac{ 2 \ \sin^{2}\theta \ \left( 2 \text{cn}^{2}\theta_{0} 
    \beta -1 \right) - v_{2} \ \sin^{2}\theta_{0} \left( \text{cn}^2 
    \theta_{0} \beta - \text{cn}^2 \theta \beta \right) }{ 
    \left( 2 \text{cn}^{2}\theta_{0} \beta  - 1 \right)
    \left\{ \left( \text{cn}^{2} \theta_{0} \beta - \text{cn}^{2} \theta
    \beta \right)^2 + \left( 2 \ \beta \text{cn} \beta \theta \ 
    \text{sn} \beta \theta \ \text{dn} \beta \theta \right)^2 
    \right\} },
							\\
  \begin{split}
    f_{2} \left( \theta, \theta_{0}, v_{2}, \beta \right) := \ & \beta \text{cn}
      \beta \theta_{0} \ \text{sn} \beta \theta_{0} \ \text{dn} \beta
      \theta_{0} + \left\{ \sin\theta_{0} \ \cos\theta_{0} \  \left(
      2 \text{cn}^{2} \theta_{0} \beta -1 \right)  - \right. 
      							\\
      & \left. - 2 \beta \text{cn}
      \beta \theta_{0} \ \text{sn} \beta \theta_{0} \ \text{dn} \beta
      \theta_{0} \ \sin^{2} \theta_{0} \right\} / v_{2} r.
  \end{split}
\end{gather*}

\noindent Equations \eqref{eq.25}-\eqref{eq.28a} are the 
solutions to the problem of a rotating gas sphere onto a Schwarzschild
black hole, i.e. they represent a relativistic generalisation of 
the accretion model first proposed by \citet{ulrich}.

\section{Convergence to known accretion models}
\label{convergence}

  We have mentioned before (cf. section~\ref{accretion-model}) that 
the analytical solution  must converge to the Ulrich accretion model
when \( \alpha \rightarrow 0 \). In order to prove this, note that three
very important  conditions are fulfilled when \( \alpha \rightarrow 0 \):
(a) the modulus \(k\) of the Jacobi elliptic functions vanishes, (b) the
root \( v_{2} \rightarrow 2 \), and (c) the parameter \(\beta \rightarrow
1 / 2 \).  These conditions together with equation~\eqref{eq.22t} imply
that relations \eqref{eq.25}-\eqref{eq.28a} naturally converge to the
non-relativistic Ulrich  model (see for example \citet{mendoza}), that is:

\begin{gather}
  r = \frac { \sin^{2} \theta_{0} }{ 1 - \cos\theta \cos\theta_{0} },
					\label{eq.29} \\
  v_{r} = -r^{ -1/2 } \left( 1 +  \frac{ \cos\theta }{ \cos \theta_{0}} 
   \right)^{ 1/2 }, 
					\label{eq.32} \\
  v_\theta = r^{ -1/2 } \frac{ \cos\theta_0 - \cos\theta }{ \sin\theta }
   \left( 1 +  \frac{ \cos \theta }{ \cos \theta_0 } \right)^{ 1/2 },
					\label{eq.31} \\
  v_\varphi = r^{ -1/2 } \frac{ \sin \theta_0 }{ \sin \theta } \left(
    1 - \frac{ \cos \theta }{ \cos \theta_0 } \right)^{ 1/2 },
					\label{eq.30} \\
  \rho =  r^{ -3/2 } \left( 1 + \frac{ \cos \theta }{ \cos \theta_0 } 
    \right)^{- 1/2 } \left\{ 1 + 2 r^{ -1 } P_2 \left( \cos \theta_{0}
    \right) \right\}^{ -1 },
					\label{eq.33} 
\end{gather}

\noindent where \( P_2(\chi) \) is the Legendre second order polynomial 
given by \( P_{2}( \chi ) := \left( 3 \cos^2 \chi - 1 \right) / 2 \).

  On the other hand, if we consider a particular case for which the 
angular momentum of the fluid particles is null, then equations
\eqref{eq.25}-\eqref{eq.28a} converge to

\begin{equation}
  v_r = - \left( 2 M / r  \right)^{ 1/2 }, 
    \qquad v_\theta = 0, \qquad 
  v_\varphi = 0, \qquad 
  n = 2^{ -1/2 } r^{ -3/2 }.
\label{eq.34}
\end{equation}

\noindent These equations  describe a radial accretion model onto a
Schwarzschild black hole. They correspond to the model first constructed
by \citet{michel72} when pressure gradients in his calculations are
negligible.

\section{The extreme hyperbolic model}
\label{ultrarelativistic}

  As mentioned in section~\ref{accretion-model}, the parameter \( \alpha \)
reaches its maximum value when \( \alpha = 1/8 \), which corresponds to a
minimum angular momentum \( h = 2 r_\text{g} \).
In this limit the module \( k \) of the Jacobi
elliptic functions is such that \( k = 1 \), \( v_2 = 4 \) and 
\( \beta = \sqrt{8} \) as can be seen from equations~\eqref{eq.18a}, 
\eqref{eq.21a} and \eqref{eq.23}.  Also, when \( k \rightarrow 1 \),
the following identities are valid \citep{lawden}:

\begin{equation}
  \text{sn} \, w \rightarrow \text{tanh} \, w, \quad 
  \text{cn} \, w \rightarrow \text{sech} \, w, \quad
  \text{dn} \, w \rightarrow \text{sech} \, w.
\label{eq.34a}
\end{equation}

 Using all these relations it follows that solutions 
\eqref{eq.25}-\eqref{eq.28a} converge to

\begin{gather}
  r = \frac{ \sin^{2} \theta_{0} \left( 2 \textrm{sech}^{2} \frac{ \sqrt{
    2 } }{ 4 } \theta_{0} -1 \right) }{ 4 \left( \textrm{sech}^{2} \frac{
    \sqrt{ 2 } }{ 4 } \theta_{0}- \textrm{sech}^{2} \frac{ \sqrt{ 2 }
    }{ 4 } \theta
   \right ) },
   						\label{eq.35}  \\
  v_{r} = - r^{ -1/2 } \frac{ \sqrt 2 }{ 2 } \ \frac{ \textrm{sech}^{2}
    \frac{ \sqrt{ 2 } }{ 4 } \theta \  \textrm{tanh} \frac{ \sqrt 2 }{
    4 } \theta }{ \sin\theta } \ f^{1/2}_{1\text{H}} \left( \theta,\,
    \theta_{0} \right),
						\label{eq.36}  \\
  v_{\theta} = r^{-1/2} \frac{ \textrm{sech}^{2} \frac{ \sqrt 2 }{
    4 } \theta_{0} - \textrm{sech}^{2} \frac{ \sqrt 2 }{ 4 } \theta }{
    \sin\theta } \ f^{1/2}_{1\text{H}} \left( \theta, \theta_{0} \right),
   						\label{eq.37}\\
 v_{\varphi} = 2 r^{-1/2} \frac{ \sin\theta_{0} }{ \sin\theta }
    \left( \frac{  \ \left( \textrm{sech}^{2} \frac{ \sqrt 2 }{ 4 }
    \theta_{0} - \textrm{sech}^{2} \frac{ \sqrt 2 }{ 4 } \theta \right)
    }{ 2 \textrm{sech}^{ 2 } \frac{ \sqrt 2 }{ 4 } \theta_{0} - 1 }
    \right)^{ 1/2 },
    						\label{eq.38}\\
 n = \frac { r^{ -3/2 } \sin\theta_{0} }{ 2 f^{1/2}_{1\text{H}} \left(\theta,
         \, \theta_{0} \right) \  f_{2\text{H}} \left( \theta, \,\theta_{0}
         \right) },
						\label{eq.39} \\
  \intertext{where}
  f_{1\text{H}} \left( \theta, \theta_{0} \right) := \frac { 2 \ \sin^{2}\theta
    \ \left( 2 \textrm{sech}^{2} \frac{ \sqrt 2 }{ 4 } \theta_{0}  -
    1 \right) - 4 \ \sin^{2} \theta_{0} \left( \textrm{sech}^{2} \frac{
    \sqrt 2 }{ 4 }\theta_{0} - \textrm{sech}^{2} \frac{ \sqrt 2 }{
    4 } \theta \right) }{ \left( 2 \textrm{sech}^{2} \frac{ \sqrt 2 }{
    4 } \theta_{0} - 1 \right) \left\{ \left( \textrm{sech}^{2} \frac{
    \sqrt 2 }{ 4 } \theta_{0} - \textrm{sech}^{2} \frac{ \sqrt 2 }{ 4
    }\theta \right)^{ 2 }+ \left( \frac{ \sqrt 2 }{ 2 } \textrm{sech}^{2}
    \frac{ \sqrt 2 }{ 4 } \theta \ \textrm{tanh} \frac{ \sqrt 2 }{ 4 }
    \theta \right)^2 \right\} },
  						\notag \\
  \begin{split}
    f_{2\text{H}} \left( \theta, \theta_{0} \right) &:= \frac{ \sqrt 2 }{ 4 }
       \textrm{sech}^{2} \frac{ \sqrt 2 }{ 4 } \theta_{0} \ \textrm{tanh}
       \frac{ \sqrt 2 }{ 4 }\theta_{0}  +  \left\{ \sin\theta_{0} \
       \cos\theta_{0} \ \left( 2 \textrm{sech}^{2} \frac{ \sqrt 2 }{ 4 }
       \theta_{0} - 1 \right) - \right.
       							\\
    & \left. - \frac{ \sqrt 2 }{ 2 } \textrm{sech}^{2} \frac{ \sqrt 2 }{ 4 }
      \theta_{0} \ \textrm{tanh} \frac{ \sqrt 2 }{ 4 } \theta_{0} \
      \sin^{2}\theta_{0} \right\} / 4 r.
      						\notag
  \end{split}
\end{gather}

  This model does not formally represent a relativistic Ulrich solution,
since the orbit followed by a particular fluid particle has
a hyperbolic Newtonian counterpart.  The solutions described by
equations~\eqref{eq.35}-\eqref{eq.39} are the exact relativistic solutions
to the numerical problem discussed by \citet{lee05} who used a
\citet{paczynsky80} pseudo--Newtonian potential.

\section{Discussion}
\label{discussion}

  \citeauthor{ulrich}'s Newtonian accretion model predicts the existence
of an accretion disc of radius \( r_\text{dN} \).  This is a natural
property of an accreting flow with rotation and has to be valid in the
relativistic case as well.  In order to see the modifications that
a full relativistic model imposes to the structure of the accretion
disc, let us start by observing what happens to a fluid particle when
it reaches the equator.  First, in the \citeauthor{ulrich} accretion
model, when any particle reaches the equator \( \theta = \pi / 2 \), it
does so at a radius \( r = h^2 / M \) according to the dimensional form
of equation~\eqref{eq.29}.  This corresponds to a stable circular orbit
about the central object only in the case of a particle with azimuthal
velocity that lies on the equatorial plane.  For the relativistic model
we have discussed so far, if this were the case, then particles
would arrive at the equator at a radius \citep{wald}

\begin{equation}
  r_\text{circ} = \frac{ h^2 }{ 2 M } \bigg\{ 1 + \left( 1 - 6 \alpha
    \right)^{1/2} \bigg\},
\label{eq.50}
\end{equation}

\noindent which corresponds to the radius of stable circular orbits.
However, when \( \theta = \pi / 2 \), equation~\eqref{eq.25} implies that
the value of \( r \) is very different from the one that would appear if
a stable circular orbit is expected according to equation~\eqref{eq.50}.
In fact, fluid particles arrive at a radius greater than \( r_\text{circ}
\).

\begin{figure}
   \begin{center}
     \includegraphics[scale=0.9]{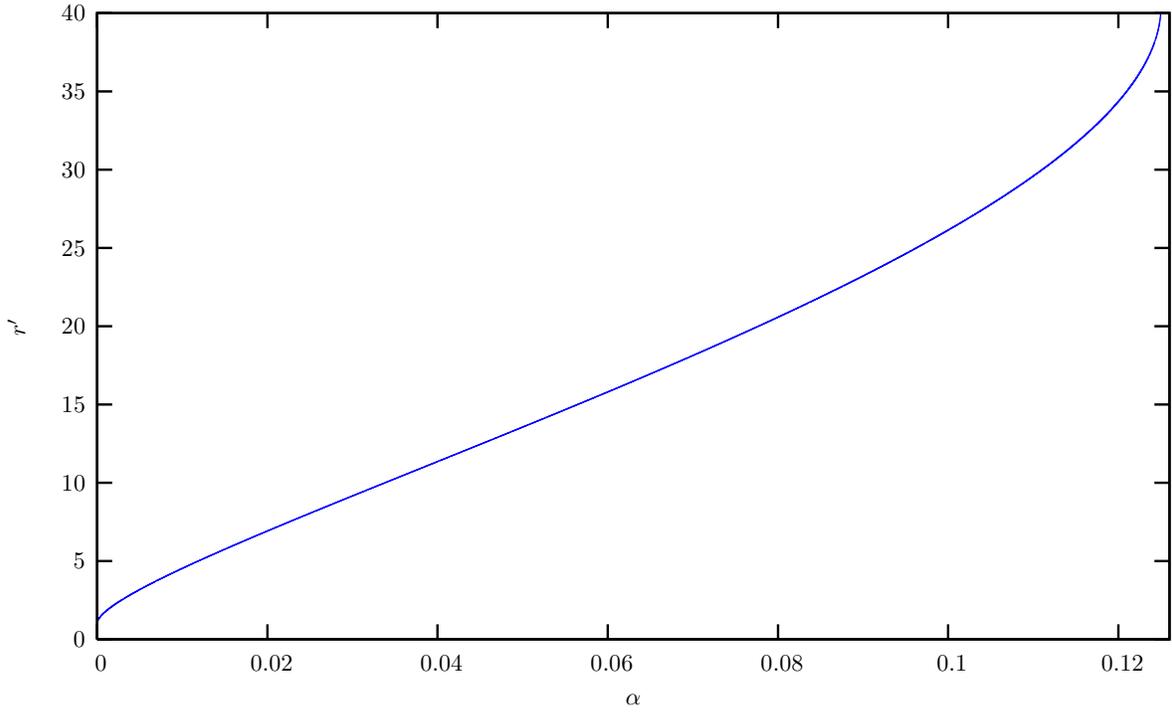}
   \end{center}
  \caption[Evolution of the radius of the disc as a function of 
           $ \alpha $ ]{ The figure shows a plot of the radius of the disc
           \( r' \) measured in arbitrary units, as a function of the 
	   parameter \( \alpha \).  In the non--relativistic case, for
	   which  \( \alpha = 0 \), the radius of the disc is exactly the
	   same as the one predicted by \citet{ulrich}.   For the
	   extreme hyperbolic model, when \( \alpha \rightarrow 1 / 8 \),
	   the radius of the disc grows without limit.} 
\label{fig01}
\end{figure}

  We can also discuss what happens to the radius of the disc \( r_\text{d}
\) for any \( \alpha \).  This radius is obtained by taking a particle
that arrives from a streamline just above the equator, i.e. \( \theta_0
= \pi / 2 - \eta \), where the positive quantity \( \eta \ll 1 \).
Figure~\ref{fig01} shows how this radius varies as a function of \(
\alpha \).  As it can be seen, the radius \( r_\text{d} \) grows
monotonically from the value \( r_\text{dN} \) when \( \alpha = 0 \)
to infinity when \( \alpha  = 1 / 8 \).   This behaviour strongly
modifies the traditional view, particularly since the disc occupies
all the equatorial plane in the extreme hyperbolic model.   The fact
that the disc radius diverges when \( \alpha = 1/8 \) can be prooved
directly using the results obtained in section~\ref{ultrarelativistic}.
Indeed, evaluating equation~\eqref{eq.35} for \( \theta = \pi / 2 \)
and then taking the limit when \( \theta_0 \rightarrow \pi / 2 \) it
follows that \( r \rightarrow \infty \).

\begin{figure}
   \begin{center}
     \includegraphics[scale=0.9]{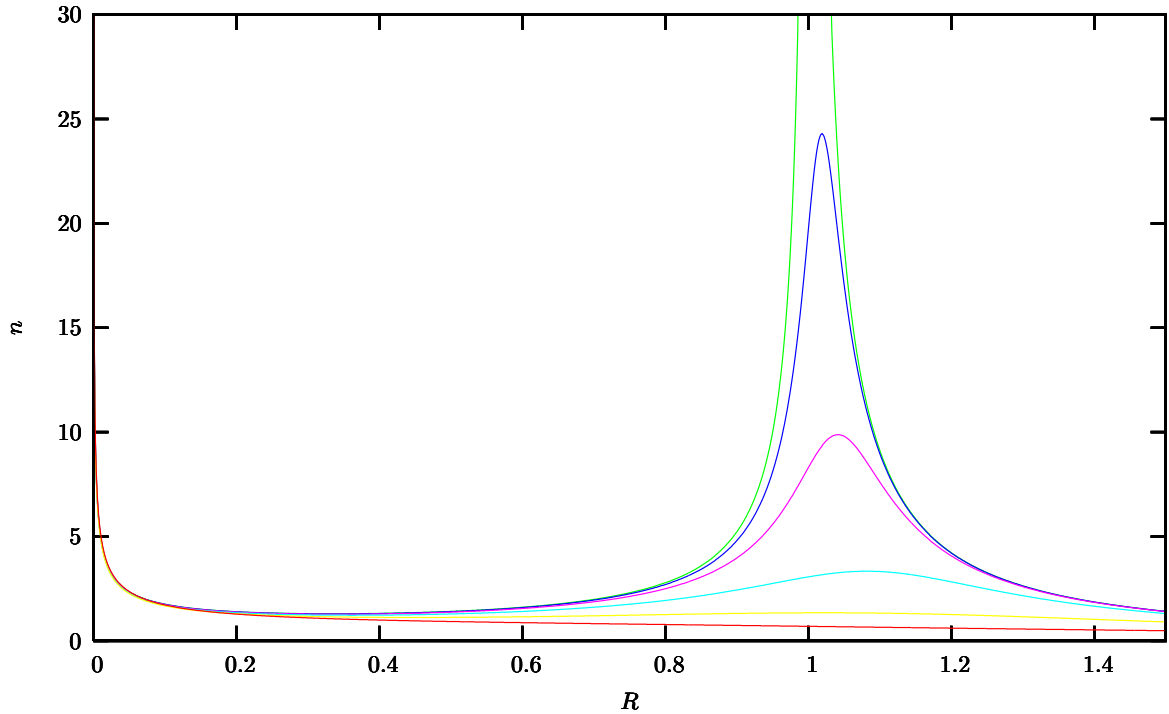}
   \end{center}
  \caption[Particle number density in equator]{
           The plots represent different particle number densities \( n \) 
	   measured in units of \( n_0
	   \), as a function of the radial distance \( R \) (measured in
	   units of \(r_{*}\)) evaluated
	   in the equator, i.e. for which the polar angle \( \theta = \pi
	   / 2 \).  From bottom to top the models correspond to values \(
	   \alpha \) of \( 1/8,\ 10^{-1},\ 10^{-2},\ 10^{-3},\ 10^{-4}
	   \) and \( 10^{-5} \).  All profiles diverge at the origin
	   because of accumulated material at that point.  The particle
	   number density diverges in the Newtonian limit (for which \(
	   \alpha \rightarrow 0 \)) at the border of the disc, which
	   corresponds for that particular case to \( R \rightarrow 1
	   \) \citep{mendoza}.  However, this singularity disappears
	   and softens the density profile in the disc as \( \alpha
	   \rightarrow 1/8 \).
	   } 
\label{fig02}
\end{figure}

  The fact that the disc radius grows monotonically as \( \alpha \)
approaches the value \( 1/8 \) means that the density of the disc
should be distributed in a more homogeneous way.  Figure~\ref{fig02}
shows density profiles evaluated in the equatorial plane \( \theta
= \pi / 2 \) as a function of the distance to the central object.
In all cases the particle number density diverges in the origin because
it represents a point of accumulated material.  The case \( \alpha =
0 \) corresponds to the non--relativistic Ulrich model and apart from
the divergence that the particle number density has at \( r = 0 \) it
also grows without limit at the radius of the disc \( r_\text{dN} \).
This is generally attributed to border effects that appear because
the disc has been assumed to be thin \citep[see e.g.][and references
therein]{mendoza}.  However, as Figure~\ref{fig02}
shows, the divergence of the particle number density at the border of
the disc disappears as soon as \( \alpha \) moves away from a null value.
Furthermore, it does so in such a way that the density of the disc varies
very smoothly throughout the disc as \( \alpha \rightarrow 1 / 8 \).

\begin{figure}
   \begin{center}
     \includegraphics[scale=0.75]{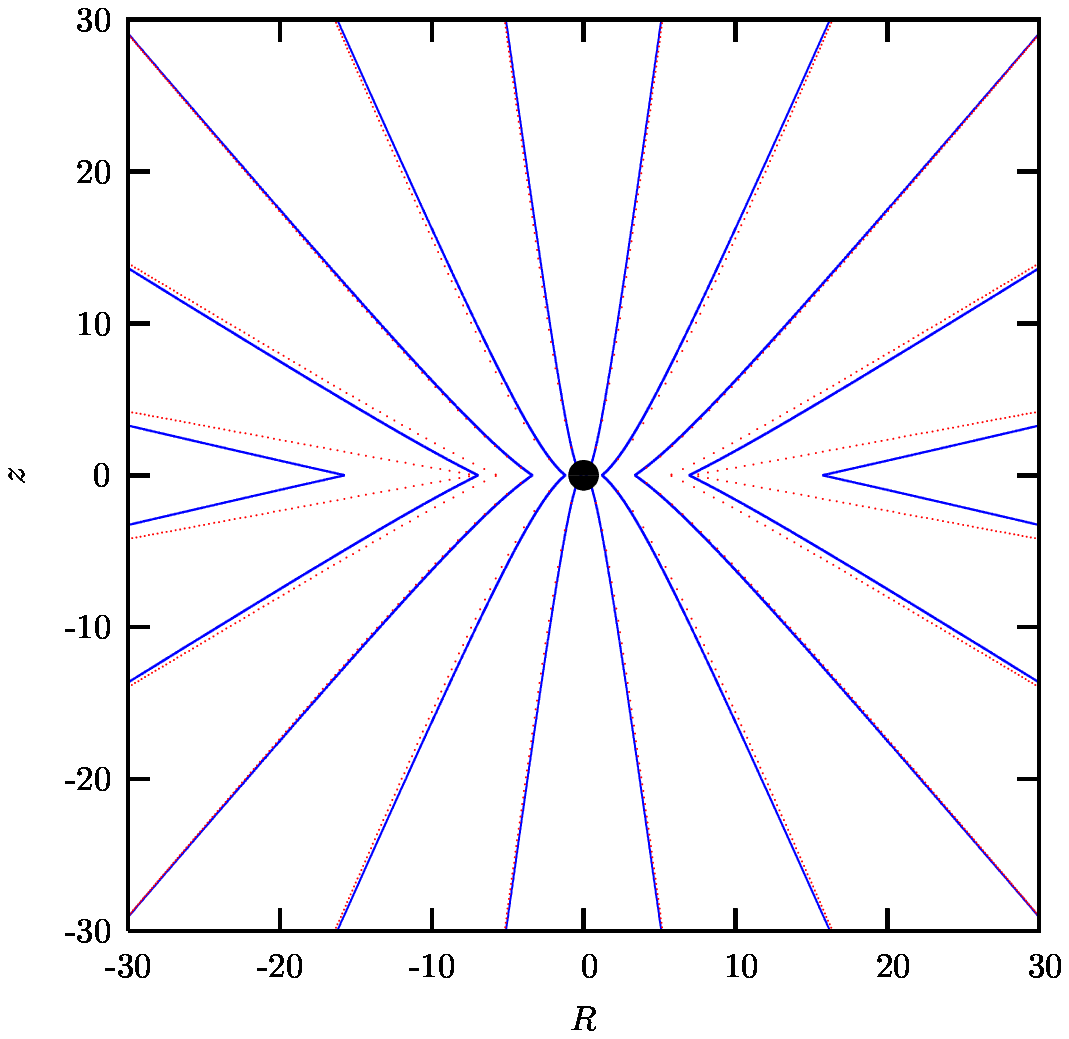}
     \includegraphics[scale=0.75]{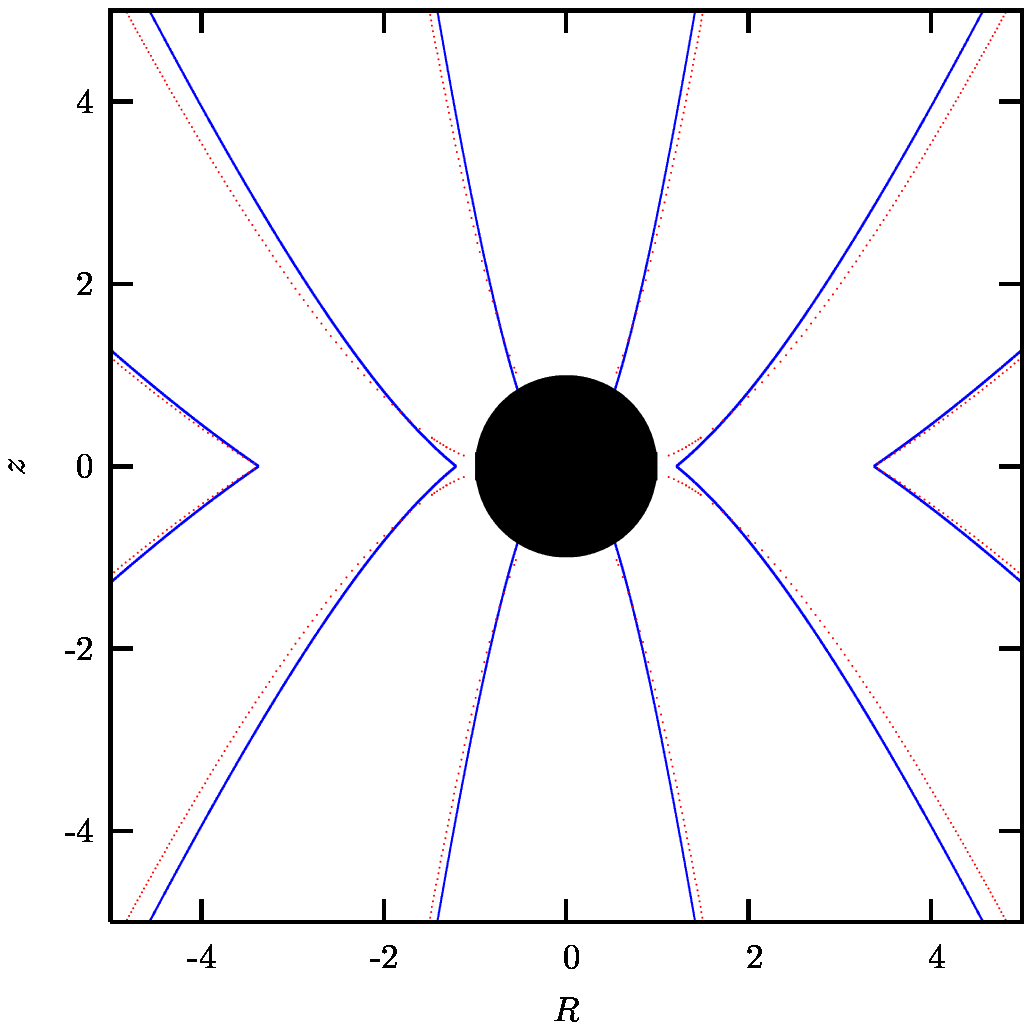}
   \end{center}
  \caption[Comparison with pseudo--Newtonian potentials]{  The figure
	   shows a comparison between the fully relativistic solution
	   presented in this article (continuous lines) with the
	   Newtonian \citet{paczynsky80} numerical approximations made
	   by \citet{lee05} (dotted lines).  Distances in the plot are
	   measured in units of the Schwarzschild radius.  The plot is
	   a projection at an azimuthal angle \( \varphi = \text{const}
	   \). The length \( R \) is the radial distance measured in
	   the equator.  In both cases, the streamlines were calculated
	   in the extreme hyperbolic case for which \( \alpha = 1 / 8
	   \), i.e. the specific angular momentum for a particular fluid
	   particle is twice the Schwarzschild radius.	Particles were
	   considered to be uniformly rotating at a distance of \( 50
	   \) Schwarzschild radius measured from the origin.  Both,
	   the small and large scale panels show that the complete
	   relativistic solution differs significantly from their
	   calculations. The pseudo--Newtonian \citeauthor{paczynsky80}
	   approximations were kindly provided by W.~H.~Lee.  }
\label{fig03}
\end{figure}

  The results of section~\ref{ultrarelativistic} can be used to compare
with the pseudo--Newtonian \citet{paczynsky80} approximation used by
\citet{lee05}.  Figure~\ref{fig03} shows a comparison between the full
relativistic solution with the pseudo--Newtonian approximation.  It is
clear from the images that the solution differs not only at small radii,
near the Schwarzschild radius, but also at large scales.  This is due to
the  strength of the gravitational field produced by the central source,
which makes particles approach the equator quite rapidly.  For instance,
near the event horizon there are fluid particles that appear to be
swallowed by the hole when described by a pseudo--Newtonian potential.
However, the complete relativistic solution shows that for this particular
case some of those particles are not swallowed directly by the hole,
but are injected to the accretion disc.

\begin{figure}
   \begin{center}
     \includegraphics[scale=0.7]{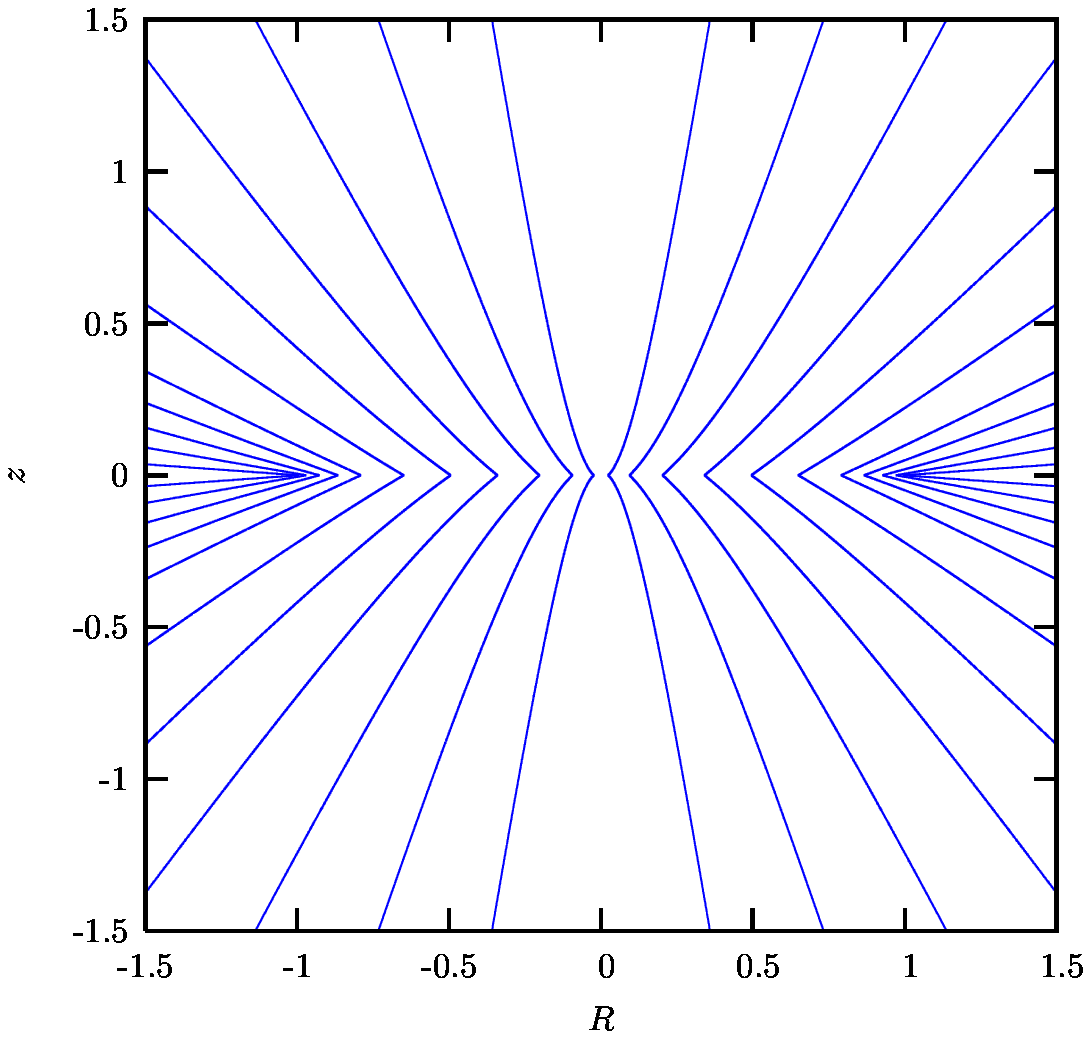}
     \includegraphics[scale=0.7]{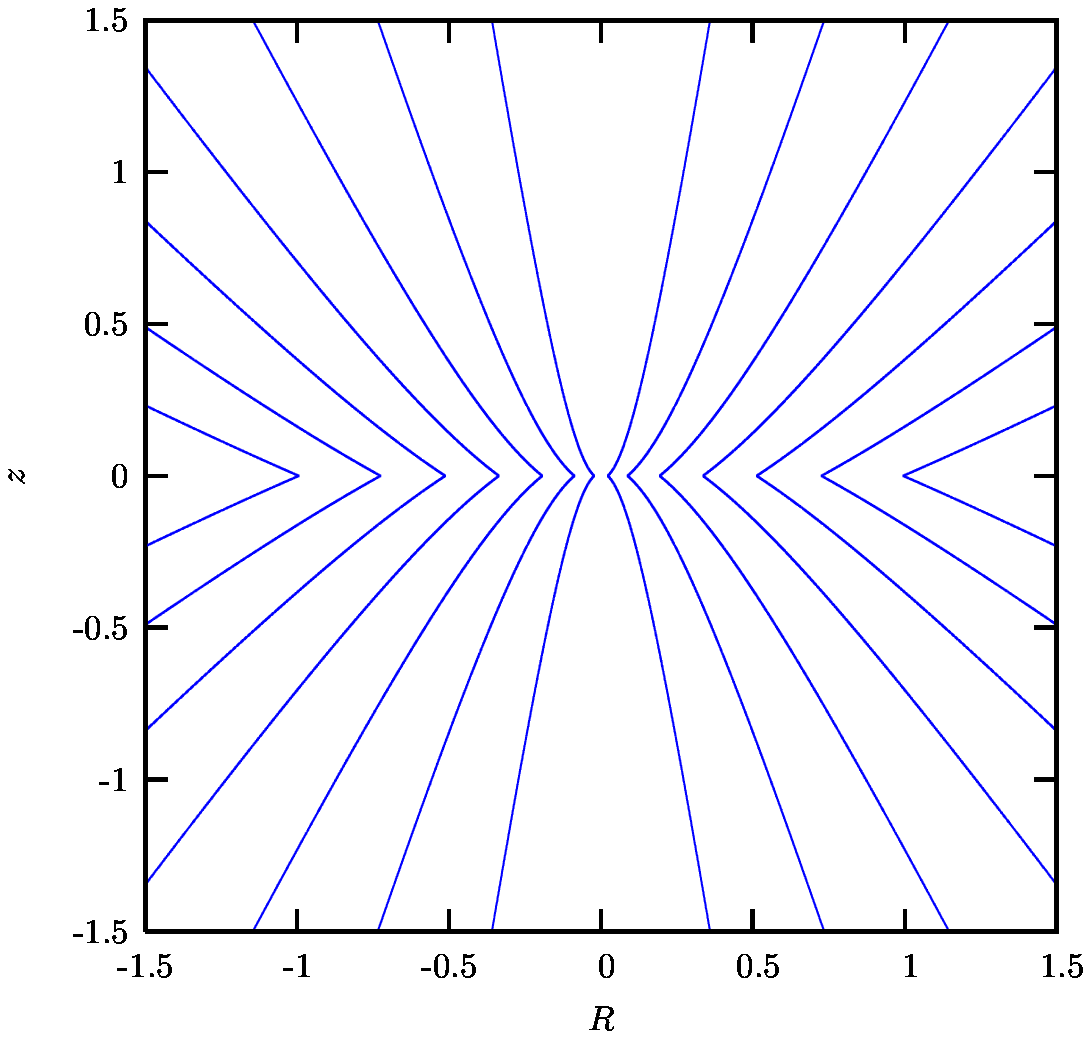}
   \end{center}
  \caption[Streamlines]{Streamlines for values of the parameter
  \( \alpha = 10^{-5},\ 0.12 \) from left to right are shown in the figure.
  Lengths are measured in units of the radius \( r_* \).  The equatorial
  radius is labelled by \( R \). The case \( \alpha = 10^{-5} \) is very
  close to the Newtonian one (see for example \citet{mendoza}).  This
  particular case shows that the streamlines are accumulated at \( R = 1
  \), which corresponds to the Newtonian radius \( r_\text{dN} \).
  However, the right panel shows that as \( \alpha  \) approaches the 
  value \( 1/8 \) the streamlines are not packed together any longer.}
\label{fig04}
\end{figure}

\begin{figure}
   \begin{center}
     \includegraphics[scale=0.75]{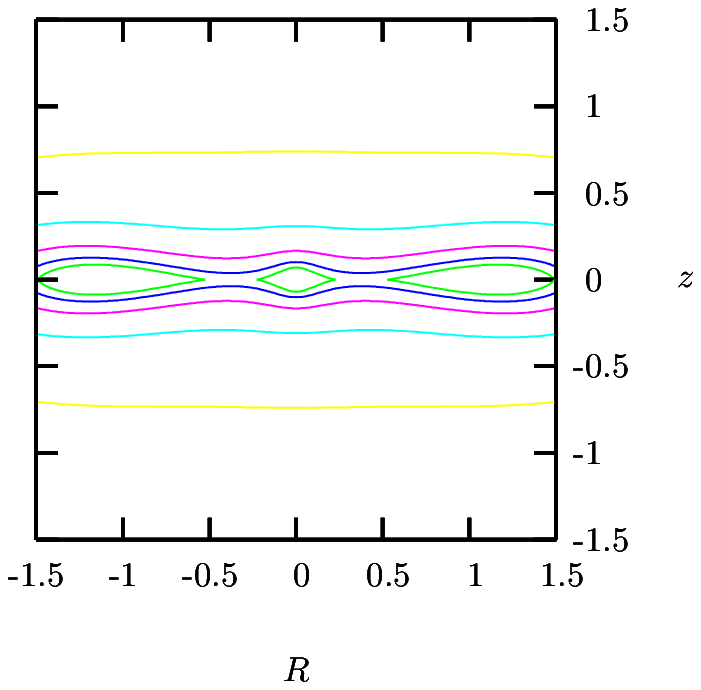}
     \includegraphics[scale=0.75]{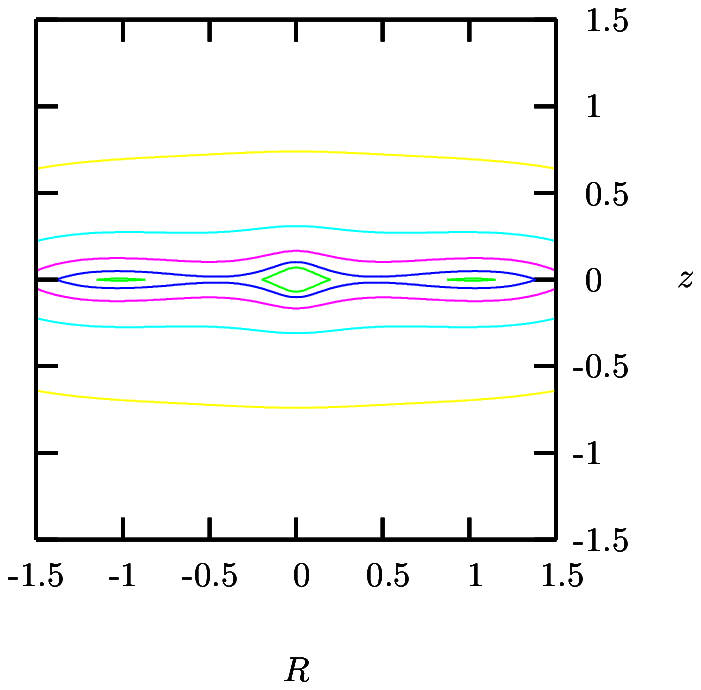}
     \includegraphics[scale=0.75]{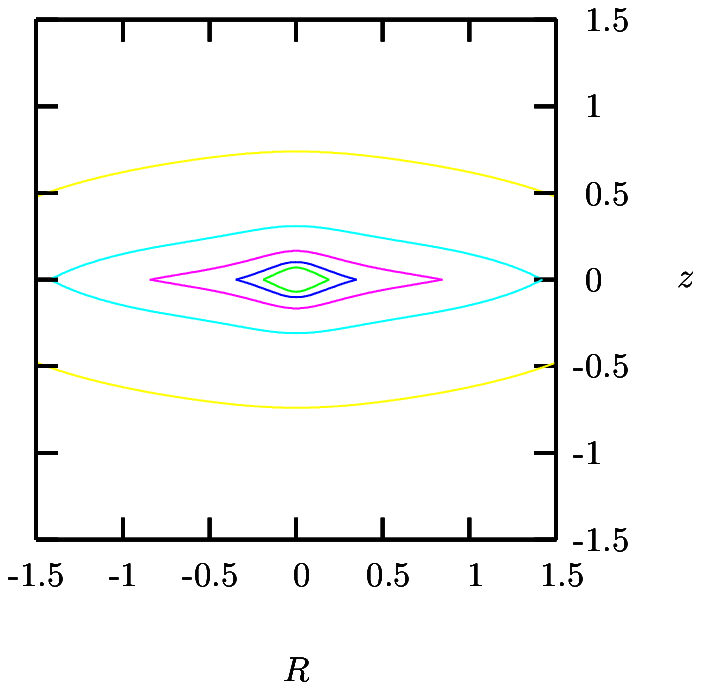}
   \end{center}
  \caption[Density isocontours]{ Particle
  number density isocontours for \( \alpha = 10^{-5},\ 0.05,\ 0.12 \)
  are shown in each diagram.  The left panel roughly corresponds to the
  non--relativistic case as described by \citet{mendoza}.  All models
  show a density divergence at the origin.  However, only the Newtonian
  case exhibits another divergence at the border of the disc \( R =
  1 \).  Lengths in the plot are measured in units of the radius \( r_*
  \) and the density isocontours correspond to values of \( n / n_0  =
  0.1,\ 0.6,\ 1.1,\ 1.6,\ 2.1,\ 2.6 \). }
\label{fig05}
\end{figure}

  The work presented in this article represents a general relativistic
approach to the Newtonian accretion flow first proposed
by \citet{ulrich}.  The main features of the accretion flow are still
valid with the important consequence that, the radius of the equatorial
accretion disc grows from its Newtonian value for the Ulrich case up
to infinity in the extreme hyperbolic situation, for which the
angular momentum is twice the Schwarzschild radius.  As a consequence,
the particle number density diverges on the border of the disc only for
the Newtonian case described by \citeauthor{ulrich}.  This is due to
the fact that, when the radius of the disc grows, the particle number
density on it rearranges in such a way that it smoothly softens as
the extreme hyperbolic case is approached.  Figures~\ref{fig04} and
\ref{fig05} show streamlines and density isocontours for different values
of the parameter \( \alpha \).

\section{Acknowledgements}
  We dedicate the present article to the vivid memory of Sir~Hermann Bondi
who pioneered the studies of spherical accretion.  We would like to
thank William Lee  for providing his numerical \citeauthor{paczynsky80}
pseudo--Newtonian results in order to make comparisons with the exact
analytic solution presented in this article.  The authors
gratefully acknowledge financial support from DGAPA--UNAM~(IN119203).

\bibliographystyle{apj}
\bibliography{acc}

\label{lastpage}
\end{document}